\documentclass[aps,prb,preprint,showpacs,superscriptaddress]{revtex4}
\usepackage{bm,amsmath,amssymb,latexsym,mathrsfs,enumerate}
\usepackage[mathcal]{euscript}
\usepackage{graphicx, subfigure}
\newcommand{\figwidth}{0.47\textwidth}

\begin{document}

\title{Large temperature dependence
of the Casimir force at the metal-insulator transition}

\author{E.~G. Galkina}
\affiliation{Advanced Science Institute, The Institute of Physical
and Chemical Research (RIKEN), Wako-shi, Saitama, 351-0198, Japan}
\affiliation{Institute of Physics, 03028 Kiev, Ukraine}
\author{ B.~A. Ivanov}
\email{bivanov@i.com.ua} \affiliation{Advanced Science Institute,
The Institute of Physical and Chemical Research (RIKEN), Wako-shi,
Saitama, 351-0198, Japan} \affiliation{Institute of Magnetism,
03142 Kiev, Ukraine} \affiliation{National T. Shevchenko
University of Kiev, 03127 Kiev, Ukraine}
\author{Sergey~Savel'ev }
\affiliation{Advanced Science Institute, The Institute of Physical
and Chemical Research (RIKEN), Wako-shi, Saitama, 351-0198, Japan}
\affiliation{Department of Physics, Loughborough University,
Loughborough LE11 3TU, UK}
\author{V. A. Yampol'skii}
\affiliation{Advanced Science Institute, The Institute of Physical
and Chemical Research (RIKEN), Wako-shi, Saitama, 351-0198, Japan}
\affiliation{A. Ya. Usikov Institute for Radiophysics and
Electronics, 61085 Kharkov, Ukraine}
\author{Franco Nori}
\affiliation{Advanced Science Institute, The Institute of Physical
and Chemical Research (RIKEN), Wako-shi, Saitama, 351-0198, Japan}
\affiliation{Department of Physics, Center for Theoretical
Physics, Applied Physics Program, Center for the Study of Complex
Systems, University of Michigan, Ann Arbor, MI 48109-1040, USA}

\date{\today}

\begin{abstract}
{The dependence of the Casimir force on material properties is
important for both future applications and to gain further insight
on its fundamental aspects. Here we derive a general theory of the
Casimir force for low-conducting compounds, or poor metals. For
distances in the micrometer range, a large variety of such
materials is described by universal equations containing a few
parameters: the effective plasma frequency $\omega_p$, dissipation
rate $\gamma$ of the free carriers, and electric permittivity
$\varepsilon_\infty$ for $\omega \geq \omega_p$ (in the infrared
range). This theory can also describe inhomogeneous composite
materials containing small regions with different conductivity.
The Casimir force for mechanical systems involving samples made
with compounds that have a metal-insulator transition shows an
abrupt large temperature dependence of the Casimir force within
the transition region, where metallic and dielectric phases
coexist.}
\end{abstract}





\maketitle

\section{Introduction and motivation}

The Casimir force~\cite{1} has demonstrated the reality of
zero-point field fluctuations, which played a significant role in
the development of quantum field theory (see, e.g., the
monographs~\cite{m1,m3} and review
papers~\cite{r1,5,r3,PhysicsToday}). The Casimir effect attracts
considerable attention because of its numerous applications in
quantum field theory, atomic physics, condensed matter physics,
gravitation and
cosmology.~\cite{m1,m3,r1,5,r3,PhysicsToday,f,BECPit,9}  The
experimental observation of the Casimir force is of fundamental
importance. Despite the fact that the magnitude of the Casimir
force is quite small, its presence is established by a number of
experiments, usually done for metallic samples; see, e.g.,
Refs.~\onlinecite{9,exp1,10,11,12,exp2}. Furthermore, this force
is relevant for various nanomechanical devices, where the space
separation of nearby plates is very small.~\cite {8,m3,r3}

\subsection{Casimir force for good metals and dielectrics}
The Casimir force between two macroscopic samples
is caused by a spatial redistribution of the fluctuations of the
electromagnetic field compared to that of free space because of
the presence of the samples. For the simplest case of two parallel
\emph{perfectly conducting} thick metallic plates placed in vacuum
and separated by a distance $l$, the Casimir force per unit area
of the sample at zero temperature can be written as
\begin{equation}
\label{eq1} F_{\mathrm{C}} =  \frac{ \pi ^2
}{240}\frac{c\hbar}{l^4},
\end{equation}
where $c$ is the speed of light, and $\hbar $ is the Planck
constant. For \emph{dielectric} bodies with frequency-dependent
dielectric permittivities, the value of this force has been found
by Lifshitz.\cite{2}  If the permittivity $\varepsilon$ is
frequency-independent, for two equivalent dielectric bodies or for
a dielectric sample and an ideal metal, this force can be written
as
\begin{equation}
\label{eq2} F_{\mathrm{L}} = \frac{ \pi ^2 }{240}\frac{c\hbar}{l^4}
\cdot \left( {\frac{\varepsilon - 1}{\varepsilon + 1}} \right)^\nu
\varphi _\nu (\varepsilon ),
\end{equation}
where $\nu = 2$ for two equivalent dielectric bodies and $\nu = 1$
for the interaction of a dielectric sample and a metal. The
function $\varphi _\nu (\varepsilon ) \to 1$ when $\varepsilon \gg
1$, and $\varphi _\nu (\varepsilon )$ decreases when $\varepsilon
\to 1$; in particular, $\varphi _1(\varepsilon \to 1) =0.46$ and
$\varphi _2(\varepsilon \to 1) =0.35$. Strictly speaking,
equations of type (\ref{eq1}) or (\ref{eq2}) are valid when $l <
\hbar c / kT$, where only the zero-point fluctuations of the
electromagnetic field are important (see Refs.~\onlinecite{3,4,5}
for details). At room temperature, this inequality is valid for
distances less then a few micrometers. Below we will only consider
this range.

\subsection{Material aspects of the Casimir force}
To study the Casimir force, different materials can be used.
Indeed, it is important to understand \emph{how} this force is
affected by the choice of different \emph{materials}. For example,
recent studies, using silicon with different degrees of
doping~\cite {semicond,Exper07} or materials for sensors, like
Vanadium oxide~\cite {Exper07}, have shown numerous specific
features which are absent in the good metals traditionally used to
study the Casimir force.

The investigation of material-dependent features of the Casimir
force is important not only for future applications, but also for
fundamental physics. To discuss the material-dependent aspects of
the Casimir force, let us note the following. The well-known
results present in the expressions (\ref{eq1}) and (\ref{eq2}) are
obtained for frequency-independent values of the electrical
permittivity $\varepsilon $. For metals, this means $\varepsilon =
\infty $ for any frequency. Detailed investigations, taking into
account the dispersion of the media, have shown~\cite{3,4} that
the universal formula of the type $F_{\mathrm{C}} \propto 1 / l^4$
is valid for distances $l > \lambda _0 $, where $\lambda _0 = c /
\omega _0 $, and $\omega _0 $ is the highest characteristic
frequency of the media. Beyond this approximation, the Casimir
force $F $ can be written~\cite{2,3,4,5} as
\begin{equation}
\label{eq3} F = \frac{\hbar }{2\pi ^2c^3} \cdot \int\limits_0^\infty
{\zeta ^3d\zeta } \cdot \Phi [\varepsilon (i \zeta)],
\end{equation}
where $\varepsilon = \varepsilon (i\zeta )$ is the complex
permittivity of the media, the summation over the Matsubara
frequencies is replaced by integration over $\zeta $ (this is
adequate~\cite{3} when $l < \hbar c / kT$), $\Phi [\varepsilon (i
\zeta)]$ is a functional of the function $\varepsilon (i\zeta )$,
\begin{eqnarray}\label{FullInt}
 \Phi [\varepsilon (i \zeta)] &=&
 \int\limits_1^\infty {p^2dp\biggl(\frac{1}{A_1^\nu e^x - 1}} +
\frac{1}{A_2^\nu e^x - 1}\biggl), \\ \nonumber
 A_1 &=& \frac{s + p}{s - p},\quad A_2 = \frac{p\varepsilon + s}{p\varepsilon
- s},\quad s = \sqrt {\varepsilon + p^2 - 1}\;,
 \end{eqnarray}
where $x=2p\zeta l/c$. Two terms in Eq.~\eqref{FullInt} describe
the contributions of the modes with two different polarizations of
the electric field, parallel to the surface and parallel to the
incidence plane (which includes the normal to the surface and the
wave vector of the photon), respectively. The exponents $\nu = 2$
and $\nu = 1$ correspond to the same cases as for Eq.~(\ref{eq2}),
namely, the interaction between two equivalent dispersive media
($\nu = 2$), and dispersive medium, interacting with an ideal
metal ($\nu = 1$). The general properties of the function
$\varepsilon (i\zeta )$ are the following: $\varepsilon (i\zeta )$
is a monotonic function of $\zeta $, and $\varepsilon (\zeta ) \to
1$ for the values of $\zeta $ higher than all the characteristic
frequencies $\zeta > \omega _0 $ of the medium. For metals, the
plasma frequency $\omega _{\mathrm p} $ is the highest frequency
$\omega _0 $. Thus, the standard Casimir result (\ref{eq1}) is
valid for large distances $l > \lambda_p = c / \omega_p $ between
the plates, see Ref.~\onlinecite{3}. For the opposite limit
case\cite{3} of smaller distances, $l < \lambda_p$,
\begin{equation}
\label{eq4} F(l \to 0) = \frac{\hbar }{8\pi ^2l^3}\bar {\omega },
\quad \bar {\omega } = \int\limits_0^\infty {\left(\frac{\varepsilon
- 1}{\varepsilon + 1}\right)^\nu d\zeta } \; ,
\end{equation}
where the real dispersion, e.g., the dependence of the media
permittivity on the frequency, is used.

\subsection{Caviats and limitations}
It is worth noting here that, as far as we know, only one
experiment \cite{10} has been performed using the parallel-plate
configuration originally envisioned by Casimir. Most measurements
of the Casimir force have studied the interaction of a spherical
probe with a flat substrate, using the so-called Proximity Force
Theorem~\cite{PrT} to relate the force for different geometries of
the experiment to the force between two parallel plates. The
experimental search for corrections to this approximation has been
done recently.~\cite{PFTcorr} For the original plane-parallel
geometry, the accuracy of the measurements~\cite{10} of the
Casimir force, done for distances ranging from 0.5 to 6
micrometers, is not very high, within 15{\%}. A significant
difficulty has been the necessity to keep the samples parallel
during the measurements at different distances. Some of these
problems, in principle, could be overcome by measuring the Casimir
force in a fixed geometry of the experiment (fixed $l$, for
plane-parallel geometry) by varying some parameters of the sample.
The media properties could be changed by varying the temperature
of the sample. Varying the carrier density of semiconductors by
laser irradiation has also been proposed
recently.~\cite{semicond,Exper07}

The Casimir force for standard metals has a weak temperature
dependence. For metals, the Drude formula, $\varepsilon = 1 +
\omega_p^2 / \zeta (\zeta + \gamma )$ is typically used, where
$\omega_p$ is the metal plasma frequency and $\gamma $ is the
relaxation rate. For typical metals like copper, aluminum or gold,
the plasma frequency is practically temperature-independent, and
the only way to modify the Casimir force by changing the metal
parameters is via the temperature dependence of $\gamma $. For
such metals, $\gamma \ll \omega_p$, and the corresponding
corrections are small. Another problem: for standard metals the
value of $\lambda_p = c / \omega_p$ lies in the ultraviolet
region, $\lambda_p \le 0.1 \ \mu$m. Thus, to observe dispersive
effects, the region $l \le \lambda _{\mathrm p} $ should be
investigated, which is quite difficult experimentally.~\cite{12}
This limitation can be overcome by using thin metallic
films,\cite{13} but even for this optimal case the temperature
corrections are not higher than a few percent.

\subsection{Casimir force for pure metals and compounds}
Numerous compounds are known for which the carrier density and
plasma frequency $\omega_p$ are abnormally small. The
investigation of such conducting systems, which can be called
``poor metals'', is of interest from the point of view of both
fundamental physics and applications. Examples include highly
doped silicon~\cite{semicond,Exper07}, left-handed
materials~\cite{left}, transition metal oxides showing the
metal-insulator transition (MIT)~\cite{MIT}, cuprate
high-temperature superconductors~\cite{HTCS}, and manganites where
the phenomenon of colossal magnetoresistance is
observed.~\cite{CMR} For all of these systems, both the free
carrier density and the plasma frequency $\omega_p$ are much
smaller than for standard good metals. This means, that in
contrast to the usual metals, $\omega_p$ is not the highest
frequency of the material. The Drude behavior is observed up to
infrared frequencies, but with a relatively large value of
$\varepsilon = \varepsilon _\infty$ when $\omega \gg \omega_p$;
this value, $\varepsilon _\infty \cong 5$--$10$, is determined by
transitions of electrons in occupied bands. Thus, $\varepsilon \ne
1$ within a wide frequency region, including the ``metallic
region'', from small $\omega$ up to a few $\omega_p$. The
dissipation rate $\gamma $ for poor metals can be quite high, of
the order of a few percent, or even a few tenths of $\omega_p$.
The manifestation of the dispersion for the frequencies
corresponding to distances of the order of a few microns provides
the possibility to control the Casimir force by varying the
parameters of the metal. Recently, measurements of the Casimir
force between a metallic sphere and a sample made with a
low-conduction medium, like silicon with different degrees of
doping and vanadium dioxide VO$_2$, were proposed~\cite{Exper07}
for small separations, around 200--400 nm.

Here we develop a general theory of the Casimir force for
low-conducting compounds, i.e., poor metals. We show that, for
distances in the sub-micrometer and micrometer ranges, the Casimir
force for a large variety of such systems can be described by
formulae that depend on a small number of parameters, without
details of the total spectral characteristics. The inhomogeneous
composite systems considered here, containing small regions of
different properties, can be described within this theory. The
application of these results to the region of the metal-insulator
transition, where the metallic and dielectric phases coexist,
produces \emph{a very pronounced temperature dependence of the
Casimir force}.

\section{Derivation of the Casimir force for poor metals}

For general dispersive media, the Casimir force is determined by
the integral in Eq.~(\ref{FullInt}). Keeping in mind the large
variety of poor-metal parameters discussed above, we now need to
develop an analytical approach to estimate the integral
(\ref{FullInt}) and to study the role of different parameters,
like $\epsilon _\infty$ or $\gamma / \omega_p$, describing the
system. Let us now use a two-scale model for $\varepsilon (\zeta
)$ as follows:
\begin{equation}\label{6}
\varepsilon = E(\zeta )\left[1 + \frac{\omega _{\mathrm p}^2 }{\zeta
(\zeta + \gamma )}\right ],
\end{equation}
where the function $E(\zeta )$ describes the high frequency
dependence of $\varepsilon $. As we will show below, the detailed
properties of this function are not important in the region of
interest: $l \sim 1$~$\mu$m. The function $E(\zeta )$ is almost
constant, $E(\zeta ) = \varepsilon _\infty $, for all the metallic
region, $\omega _{\mathrm p} \sim \omega \ll \omega _0 $, and
tends to one for $\omega \gg \omega _0 $. Obviously, for such a
model the standard Casimir behavior in Eq.~(\ref{eq1}) is valid at
large enough distances: $l \gg \lambda_p \sim 1 \ \mu $m.

To calculate the Casimir force for distances of the order of
$c/\omega_{\mathrm p} \sim 1$~$\mu$m we use the general equation
(\ref{eq3}) rewritten as
$$
F = \frac{\hbar }{2\pi ^2c^3}\left[\int\limits_0^{\left\langle
\omega \right\rangle} \zeta^3d\zeta \Phi\left[\varepsilon_\infty
\cdot \left(1 + \frac{\omega _{\mathrm p}^2 }{\zeta (\zeta+ \gamma
)}\right ) \right] \right.
$$
$$
\left. + \int\limits_{\left\langle \omega \right\rangle }^\infty
{\zeta ^3d\zeta \Phi [E(\zeta)]} \right] .$$ Here the value
$\left\langle \omega \right\rangle $ is chosen in the intermediate
region:
$$
 \omega_p \ll \left\langle \omega \right\rangle \ll \omega
_0 .$$ Therefore, we replaced $E(\zeta)$ by $\varepsilon_\infty$
in Eq.~(\ref{6}) for the first integral and omitted the Drude
multiplier for the second integral.

Expanding the integration region over $\zeta $ in both integrals
up to $0 \le \zeta < \infty $, and subtracting the extra terms, we
present the Casimir force in the form,
\begin{equation}
\label{eq5} F = F^{(m)} + \Delta F,
\end{equation}
with
\begin{equation}
\label{Fm} F^{(m)}=\frac{\hbar }{2\pi ^2c^3}\int\limits_{0}^\infty
{\zeta ^3d\zeta \Phi \left[\varepsilon_\infty\left(1 +
\frac{\omega_p^2 }{\zeta (\zeta + \gamma )}\right ) \right]},
\end{equation}
\begin{equation}
\label{DeltaFf} \Delta F =\frac{\hbar }{2\pi
^2c^3}\int\limits_{0}^\infty {\zeta ^3d\zeta \biggl[\Phi
\left[E(\zeta)\right]-\Phi[\varepsilon_\infty]\biggr]}.
\end{equation}

In the frequency region, $\omega \sim c/l$, which is an important
regime for $\Phi [\varepsilon (\zeta)]$, the functions $\Phi
[E(\zeta)]$ and $\Phi [\varepsilon_\infty]$ in Eq.~(\ref{DeltaFf})
almost cancel each other. Therefore, the term $\Delta F$ is
relatively small. A more detailed analysis gives
$$\Delta F \, \cong \, \frac{\hbar c\lambda_0^2}{l^6} \, \ll \, F.$$
Thus, in the region of interest, $\lambda _0 \ll l\sim \lambda_p$,
the Casimir force is described by the first term in
Eq.~(\ref{eq5}), $F = F^{(m)}$.

Now we introduce the variable $z = \zeta l / c$ and write the main
contribution to the Casimir force as
\begin{equation}\label{PiIntr}
F = \frac{ \pi ^2 }{240}\frac{c\hbar}{l^4}\cdot \Pi \equiv
F_{\mathrm{C}} \cdot\Pi \;,
\end{equation}
where $F_{\mathrm{C}}$ is the Casimir force \eqref{eq1} for ideal
metals, the prefactor $\Pi $ depends only on the dimensionless
parameters $\tilde {l} = l / \lambda_p$, $\varepsilon_\infty $,
and $\alpha = \gamma / \omega_p$,
\[
\Pi = \frac{120}{\pi ^4}\int\limits_0^\infty {z^3}
dz\int\limits_1^\infty p^2dp
\]
\begin{equation}
\label{eq7} \times \left[ \frac{1}{A_1^\nu \exp (x) -
1}+\frac{1}{A_2^\nu \exp (x) - 1} \right],
\end{equation}
where  $A_{1} $ and $A_{2} $ are given by Eq.~\eqref{FullInt} with
$$\varepsilon = \varepsilon _\infty \left[1 + \frac{\tilde {l}^2}{z(z +
\alpha \tilde {l})}  \right].$$

\subsection{Computing the Casimir force integrals}

To proceed further, let us change the variable $z$ by $x=z/2p$ in
Eq.~(\ref{eq7}), and note, that the integral $\int\limits_0^\infty
{ (Ae^x - 1)^{ - 1}x^3dx} $, with $A = A(x,p)$ being a smooth
function of $x$, can be approximated by $1 / A(x_0 ,p)$ with $x_0
\sim 4$. The problem is then reduced to calculating two
one-dimensional integrals, $J_1$ and $J_2$:
\begin{equation}
\label{eq8} J_{1,2} = \int\limits_1^\infty {\frac{dp}{p^2}}
\frac{1}{A_{1,2}}, \quad \Pi = \frac{1}{2}(J_1+J_2),
\end{equation}
where $A_{1}$ and  $A_{2}$ are given by Eq.~(4) with the
substitution
\begin{equation}
\label{eq9} \varepsilon = \varepsilon _\infty \left[1 +
\frac{4\tilde {l}^2p^2}{x_0 (x_0 + 2p\alpha \tilde {l})}\right].
\end{equation}
The validity of this approximation is confirmed by the numerical
calculation of integral (\ref{eq7}), as shown in Fig.~1. The
function $\Pi(\tilde {l} )$ found numerically is shown in
Fig.~\ref{fig2}. A simple analysis of Eq.~\eqref{eq8} gives us two
limit cases.

\begin{figure}[!tb]
\includegraphics[width=\figwidth]{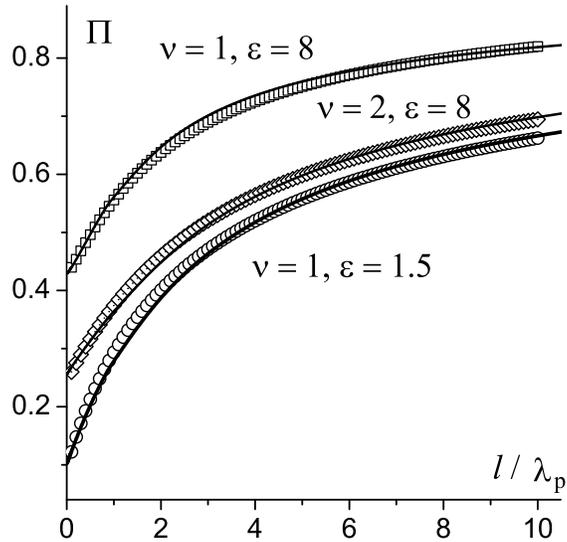}
\caption{\label{fig1} The normalized Casimir force
$\Pi=F/F_{\mathrm{C}}$ versus $l/\lambda_p$ for some values of
parameters (shown near curves). This plot compares the results of
numerical calculations of the two-dimensional integral
Eq.~(\ref{eq7}) for the prefactor $\Pi$ (symbols) with the results
within the approximate approach based on Eq.~(\ref{eq8}) (solid
curves).  This plot shows a very good agreement between both.}
\end{figure}
\begin{figure}[!tb]
\includegraphics[width=\figwidth]{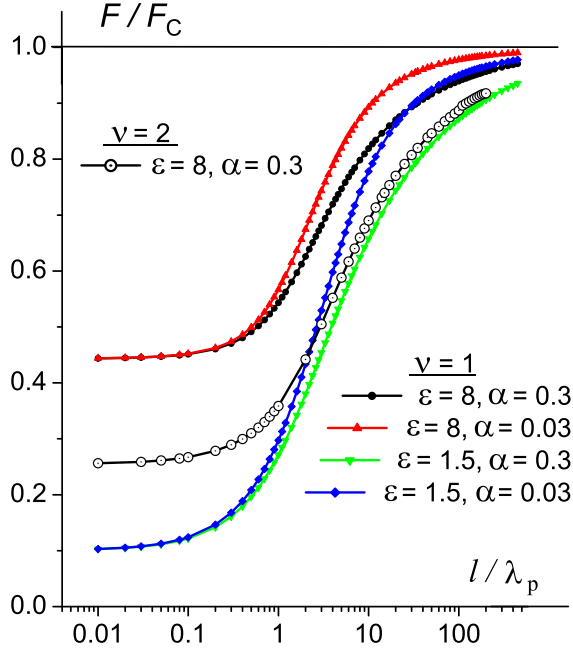}
\caption{\label{fig2} (Color online) The normalized Casimir force
$\Pi=F/F_{\mathrm{C}}$ versus the parameter $\tilde {l} = l /
\lambda_p$, for $\nu = 1,2$, using the typical value $\varepsilon
_\infty = 8$, as well as the smaller $\varepsilon _\infty = 1.5$,
and different values of the dissipation parameter $\alpha =
\gamma/\omega_p$. The horizontal line on top gives the asymptotic
value $F/F_{\mathrm{C}}=\Pi =1$ for an ideal metal.}
\end{figure}

For small $\tilde {l} \leq 0.3  $ the value of $\alpha$ plays a
minor role. In this region, the value of $\Pi$ practically does
not depend on $\tilde {l}$ and reproduces well the Lifshitz's
result \eqref{eq2} for dielectric media with a $\zeta
$-independent $\varepsilon = \varepsilon _\infty $ and $\gamma =
0$,
$$
 \Pi_{\mathrm{L}}\equiv \left(\frac{\varepsilon_\infty -
1}{\varepsilon_\infty + 1}\right)^\nu \cdot \varphi _\nu
(\varepsilon_\infty).
$$
We now emphasize that the dependence of the Casimir force,
proportional to $\bar {\omega } / l^3$, see Eq.~\eqref{eq4},
\emph{is not realized} for any $\varepsilon_\infty \ne 1$.

Otherwise, in the limit case $\tilde {l} \to \infty $, the
integrals $J_1=J_2=1$, and the ideal Casimir limit (\ref{eq1}) is
recovered. In contrast to the case of small values of  $\tilde {l}
$, the dependence of $\Pi$ on $l$ for large, but finite values of
$\tilde {l} $ shows an interesting and unexpected behavior: the
approach to saturation is quite slow, especially for large values
of
$$\alpha=\frac{\gamma}{\omega_p}\,.$$ In other words, it is hard to
reach the metallic limit value of $\Pi =1$ when $\alpha > 0.1$,
for the most interesting region $\tilde {l} \leq 10$.

\begin{figure}[!tb]
\includegraphics[width=\figwidth]{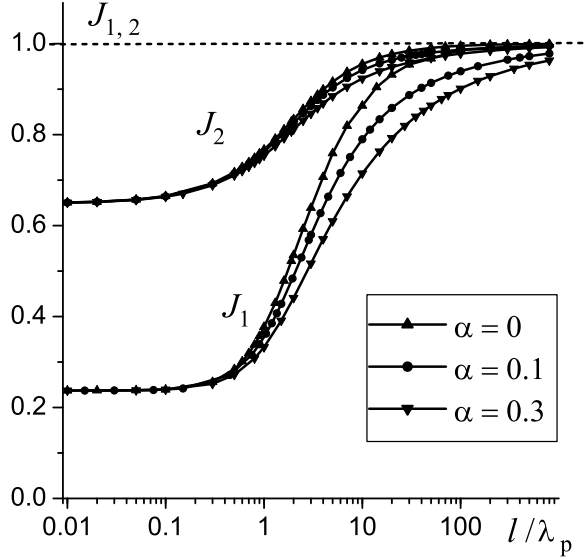}
\caption{\label{J1andJ2} Integrals $J_1$ and $J_2$, defined by
Eq.~(\ref{eq8}), describing the Casimir force versus $l/\lambda_p$
for $\varepsilon _\infty = 8$ and different values of the
dissipation parameter $\alpha = \gamma/\omega_p$ (shown in the
figure). Symbols depict the results of numerical calculations. }
\end{figure}

To understand this behavior, let us now investigate in more
details the factor $\Pi$ for not so small values of $\tilde {l}$.
As has been mentioned above, it is a sum of two contributions from
the electromagnetic fields of different polarizations. It is
convenient to examine the first and second integrals separately.
Numerical calculations show that the behavior of these two
integrals, $J_1$ and $J_2$, is essentially different for the same
values of parameters, as shown in Fig.~\ref{J1andJ2}.

\begin{figure}[!tb]
\includegraphics[width=\figwidth]{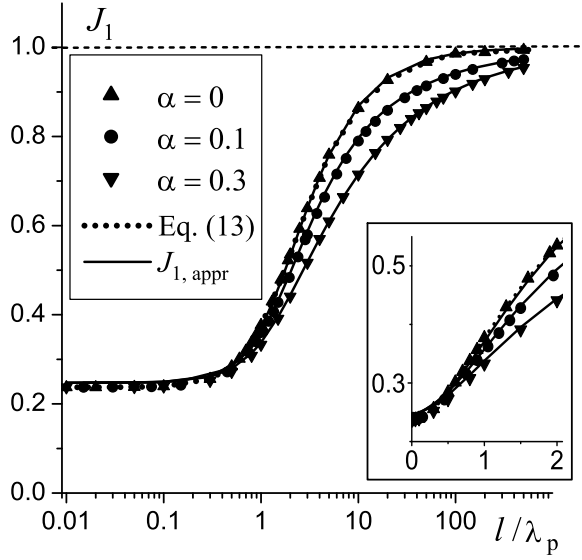}
\caption{\label{figJ1} The integral $J_1$ versus $l/\lambda_p$ for
$\varepsilon _\infty = 8$ and different values of the dissipation
parameter $\alpha = \gamma/\omega_p$. Symbols depict the results
of numerical calculations. The detailed behavior of $J_1$ for
small $\tilde{l}$ is present in the inset. The dotted line
describes the analytical result Eq.~\eqref{J1analyt} for $\alpha =
0$, solid lines are drawn in accordance with the approximate
formula Eq.~\eqref{J1}.}
\end{figure}

The two interesting features (i.e., the slow approach to
saturation and the essential dependence of $\Pi$ on $\alpha=\gamma
/\omega_p$) are mostly associated with the first integral, $J_1$,
which describes the contribution of the fluctuations with the
electric field parallel to the surfaces of plane-parallel samples.
This integral $J_1$ can be calculated analytically. For $\alpha =
0 $, it can be written as
\begin{multline}\label{J1analyt}
   J_1 = 1-\frac{2}{b}\cdot \ln \left(
   \frac{\sqrt{a^2+b^2}+b}{a}\right)+  \\
   +\frac{4}{b\sqrt{a^2-1}}\cdot\arctan \left( \frac{\sqrt{a^2-1}}{a+1}
   \cdot \frac{\sqrt{a^2+b^2}+b-a}{\sqrt{a^2+b^2}+b+a}\right) \;,
\end{multline}
where we introduce the notation
$$
 a^2=1+\varepsilon_{\infty}\frac{\tilde{l}^2}{4}, \quad
 b^2=\varepsilon_{\infty}-1.
$$
For non-zero dissipation rate, $\alpha \neq 0 $, the analytical
formula for $J_1$ is very long and inconvenient for real
estimates. But for the case of interest, $\tilde {l} \geq 3$, the
integral $J_1$ can be well approximated with the simpler
expression
\begin{equation}\label{J1}
J_{1, \, \mathrm{appr}}=\frac{\sqrt{(3+
\varepsilon_{\infty})(3+5\alpha \tilde {l}) +3
\varepsilon_{\infty}\tilde{l}^2 }-2\sqrt{3+5\alpha \tilde
{l}}}{\sqrt{(3+ \varepsilon_{\infty})(3+5\alpha \tilde {l}) +3
\varepsilon_{\infty}\tilde{l}^2 }+2\sqrt{3+5\alpha \tilde {l}}}\;,
\end{equation}
as shown in Fig.~\ref{figJ1}

This equation explains the complicated behavior of the factor
$\Pi=F/F_{\mathrm{C}}$ for large $\tilde {l}$, and especially, the
role of the dissipation constant $\alpha$. For any finite value of
$\alpha$ and extremely large $\tilde {l}$, $\tilde {l} \gg 1$ and
$\tilde {l} \gg 1/\alpha$, $J_1$ versus $\tilde {l}$ has a very
slow inverse-square-root dependence,
\begin{equation}\label{J1sqrt}
J_{1} \simeq 1- \frac{4\sqrt{5 \alpha}}{\sqrt{3
\varepsilon_{\infty}}} \cdot \frac{1}{\sqrt{\tilde{l}}} , \quad
\tilde {l} \gg 1, \, 1/\alpha .
\end{equation}
For very small $\alpha \ll 1$, the intermediate region $1/\alpha \gg
\tilde {l} \gg 1 $ appears. For this region, the behavior of $J_1$
is sharper,
\begin{equation}\label{J1oneToL}
J_{1} \simeq 1- \frac{2}{\sqrt{\varepsilon_{\infty}}}\cdot
\frac{1}{\tilde{l}} \, , \quad 1/\alpha \gg \tilde {l} \gg 1\;.
\end{equation}

\begin{figure}[!tb]
\includegraphics[width=\figwidth]{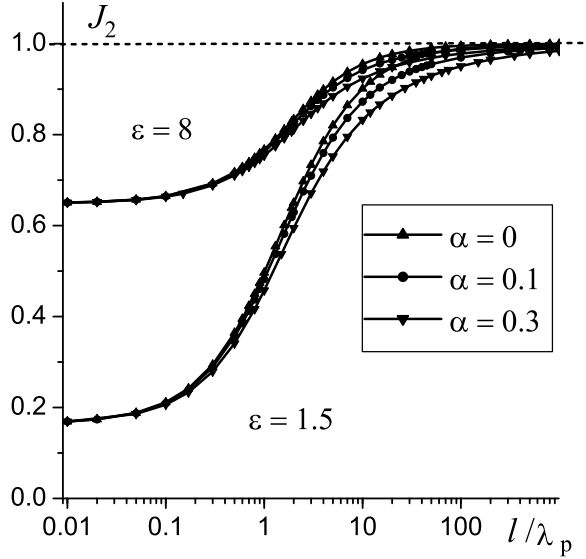}
\caption{\label{figJ2numer}  The integral $J_2$  calculated
numerically for $\varepsilon _\infty = 8$ and $\varepsilon _\infty
= 1.5$ and different values of $\alpha $ (symbols). The shapes of
the curves depend very weakly on $\alpha $.}
\end{figure}

An analytical expression for $J_{2}$ in terms of elementary
functions cannot be written. Fortunately, the integral $J_{2}$
exhibits a simpler behavior than $J_{1}$, and for its description
we can use a simple approximation. First note the quite weak
dependence of the shape of the function $J_{2}(\tilde{l})$ on the
value of $\alpha$, as shown in Fig.~\ref{figJ2numer}. For the
regime of interest here, $\varepsilon_{\infty} \gg 1$, the
difference between the values of $J_{2}$ for $\alpha =0.3 $ and
$\alpha =0 $ is maximum near the range $\tilde {l} \sim $
(10--15), and does not exceed 3{\%}. Indeed, all the curves with
$0 \leq\alpha < 0.3 $ merge together, and for describing of
$J_{2}$ within an accuracy of 1.5{\%}, the function $J_{2}(\tilde
{l})$ found for $\alpha =0.1 $ can be used. Even for small
$\varepsilon_{\infty} = 1.5$, the inaccuracy of this approximation
is less than 5{\%}.

\begin{figure}[!tb]
\includegraphics[width=\figwidth]{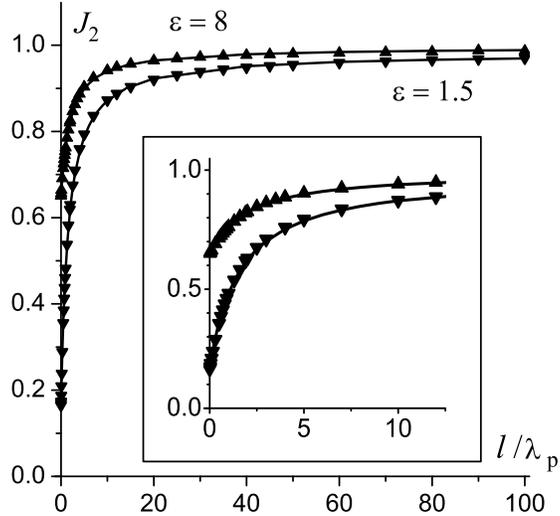}
\caption{\label{figJ2fit} The integral $J_2$ calculated
numerically for $\alpha = 0.1$ and very different values of the
dielectric permittivity, $\varepsilon _\infty = 8$ and
$\varepsilon _\infty = 1.5$ (symbols). The curves describe the
approximating function Eq.~\eqref{J2fit}. Note the very good
agreement between numerical data and approximating functions.}
\end{figure}

Numerical data are well fitted by the very simple formula
\begin{equation}\label{J2fit}
J_{2, \, \mathrm{fit}}=\frac{J_{2, \, \mathrm{L}}+ \eta \, \tilde
{l}}{1+\eta \, \tilde {l}} \;,
\end{equation}
where $\eta \simeq $ (0.5--0.6), $J_{2, \, \mathrm{L}}$ determines
the value of $J_2$ for small $\tilde {l} \ll 1$, as shown in
Fig.~\ref{figJ2fit}. The quantity $J_{2, \, \mathrm{L}}$ describes
the contribution of $J_2$ to the Lifshitz's result \eqref{eq2} for
dielectric media with $\varepsilon = \varepsilon _\infty $ and
$\alpha = 0$.

Thus, we can present a simple description of the second integral
$J_2$: it is practically independent  on the dissipation parameter
$\alpha $, and the dependence on $\varepsilon_{\infty}$ is
governed only by the Lifshitz contribution $J_{2, \, \mathrm{L}}$.
The asymptotic behavior of $J_2$ at large distances $\tilde{l} \gg
1$ is of the form $1-(1-J_{2, \, \mathrm{L}})/\eta \tilde{l} $,
which is much weaker than the inverse-square-root
dependence~\eqref{J1sqrt} for $J_1$. For large
$\varepsilon_{\infty} \gg 1$, when $(1-J_{2, \, \mathrm{L}})
\propto 1/\varepsilon_{\infty} \ll 1$, the dependence
$J_2(\tilde{l})$ is especially weak, even compared with that for
$J_1$ in the intermediate region \eqref{J1oneToL}.

\subsection{Change in the Casimir force near the metal-insulator
transition}

The analytical formulae derived above give a good description of
the behavior of the Casimir force when the metal-insulator
transition occurs. Usually, the metal-insulator transition is
associated with an abrupt change, by a few orders of magnitude, of
the conductivity at a transition temperature $T = T_c$. Let us
start with a rough picture, assuming that a metallic phase has a
finite value of the plasma frequency whereas for the dielectric
phase the plasma frequency is zero. Using the results obtained
above, one can expect a drastic change of the Casimir force
between two plane-parallel samples caused by the change of the
parameter $\lambda _{\mathrm p} $, very near the metal-insulator
transition.

We stress that the change of the force is not connected with
changing the \textit{physical} distance $l$, but with changing the
\emph{dimensionless} quantity $\tilde {l} = l / \lambda_p$, caused
by the change of the plasma wavelength $\lambda _{\mathrm p}
=c/\omega_p$. Thus, one can expect a jump-like behavior of the
Casimir force when changing the temperature across $T_c$. Within
the transition region, the force changes from the ``metallic''
value $F^{>}$, typical for finite values of $\tilde {l}$, to the
very different value $F^{<}$, for an insulator when $\tilde {l}
\ll 1$.

The important quantity here is the change of the Casimir force,
$\Delta F=(F^{>}-F^{<})$. To estimate $F^{<}$, we can use the
Lifshitz formula \eqref{eq2} valid in the limit $l \ll \lambda_p$,
which corresponds to the dielectric phase. The value of $F^{>}$ in
the metallic phase corresponds to large, but finite values of $l
/\lambda_p$. To estimate $F^{>}$, note that the dependence of
$\Pi$ on $l$ at $10 \gtrsim l /\lambda_p \gtrsim 2.5$ is mainly
provided by the integral $J_1$,  whereas $J_2$ can be replaced by
one. Thus, the concrete value of the coefficient $\eta$ in
Eq.~\eqref{J2fit} is not important. Combining all these data
together, and restoring the initial parameters of the media,
$\omega_p$ and $\gamma$, we arrive at the simple estimate,
\begin{equation}
\label{DeltaF} \Delta F = \frac{ \pi ^2 }{240}\frac{c\hbar}{l^4}
\cdot\left[1- \left( {\frac{\varepsilon_\infty -
1}{\varepsilon_\infty  + 1}} \right)\varphi _1(\varepsilon_\infty
) - \frac{2}{\omega_p}\sqrt{\frac{5 c \gamma}{3 \varepsilon_\infty
l}} \right],
\end{equation}
where the function $\varphi _1 (\varepsilon )$ describes the
Lifshitz's result for the interaction of a dielectric sample and
an ideal metal. When the value of $\gamma$ is small enough, as for
manganites, for the distance $l$ such that $l \ll c/\gamma$,
Eq.~(\ref{J1oneToL}) is valid, and the formula for $\Delta F $
reads
\begin{equation}
\label{DeltaF2} \Delta F = \frac{ \pi ^2 }{240}\frac{c\hbar}{l^4}
\cdot\left[1- \left( {\frac{\varepsilon_\infty -
1}{\varepsilon_\infty  + 1}} \right)\varphi _1(\varepsilon_\infty
) - \sqrt{\frac{ c }{\omega_p \varepsilon_\infty \, l}} \right].
\end{equation}

Note that our results differ significantly from the theoretical
estimates given in Ref.~\onlinecite{Exper07}. In particular, the
value of $\Delta F$ in Ref.~\onlinecite{Exper07} is proportional
to the temperature $T$. The linear dependence $\Delta F$ on $T$
can be expected for large enough separations $l \gg c \hbar /kT $,
that is, larger than a few microns, and cannot appear for small
separations.

It is worth noting that the relative change of the force $$\Delta
F=\frac{F^{>}-F^{<}}{F^{>}}$$ is larger for long distances $l$,
when the value $F^{>}$ of the force for media in the conducting
state is larger than the limit value $F^{<}$ describing the case
of small $\omega_p$ and small $l/\lambda_p $. This feature is
determined by the quite slow change of the function $\Pi
(\tilde{l})$ at not so small values of $\varepsilon_\infty$, as
shown in Fig.~\ref{fig2}.

\section{Composite media and the intermediate region for
metal-insulator transition}

The very abrupt (by a few orders of magnitude) change of the
conductivity at the metal-insulator transition occurs for the dc
case only. At finite frequencies, the behavior of the complex
permittivity of compounds near metal-insulator transition is more
complicated and the jump-like behavior, typical for the static
conductivity, does not arise for $\varepsilon = \varepsilon
(i\zeta )$. Within the finite transition region, the presence of a
non-uniform state with coexisting metallic and insulator phases is
well established for all systems showing a metal-insulator
transition. Obviously, this effect is of great interest for
studying the Casimir force. The effective-medium approach suggests
that the metallic and insulating regions coexist as
interpenetrating clusters, providing a percolation
picture~\cite{EffMedia} of the metal-insulator transition at $T =
T_c $. When the transition is of first order, the phase-separated
regions are mesoscopic, in the 100 nanometer range, and
quasistatic objects (giant clusters) have approximately equal
electron densities.

\begin{figure}[!tb]
\includegraphics[width=\figwidth]{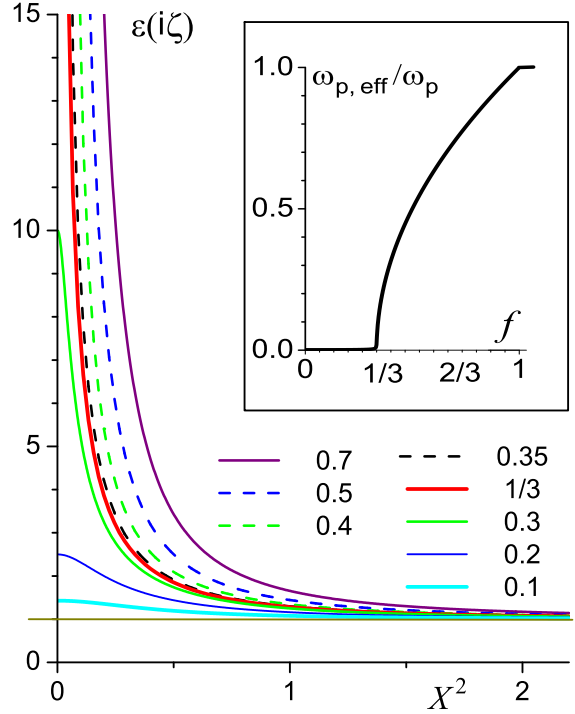}
\caption{\label{fig3ab} (Color online) The effective permittivity
$\varepsilon _{\mathrm{eff}} (i\zeta )$ (in units of $\varepsilon
_\infty )$, for $d = 3$ and different concentrations $f$ of the
metal phase, as a function of the dimensionless variable $ X^2 =
\zeta (\zeta + \gamma ) / \omega_p^2 $. Inset: effective plasma
frequency $\omega_{p,\,\mathrm{eff}} $ (in units of $\omega_p)$
versus $f$, in the coexistence region.}
\end{figure}

\begin{figure}[!tb]
\includegraphics[width=\figwidth]{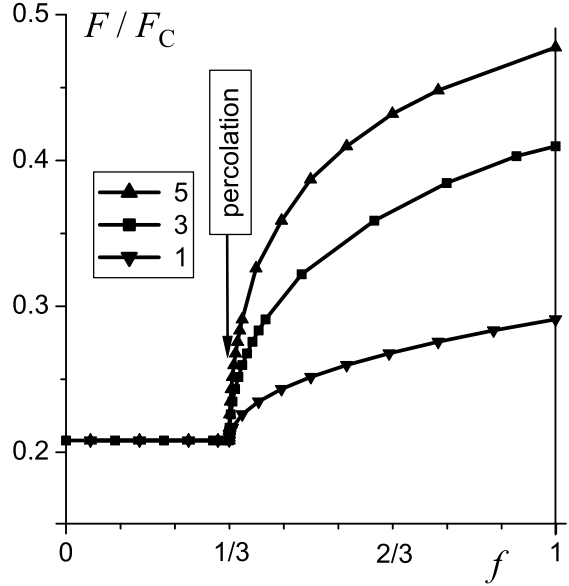}
\caption{\label{fig4} Casimir force for the interaction between
two equivalent poor metals ($\nu = 2)$ (with $\varepsilon _\infty
= 8$ and $\alpha = 0.3$) versus the concentration $f$ of the
metallic phase in the coexistence region. The force is normalized
by values of the force $F_{\mathrm{C}}$ for ideal metals in the
same geometry, and for different values of $l / \lambda_p$, where
$\lambda_p $ is determined by the plasma frequency in the pure
metallic phase.}
\end{figure}

To describe the Casimir force for such a nonuniform state, we have
used the effective-medium approximation,~\cite{EffMedia} developed
for composite metal-insulator media. This approximation has been
used for explaining the optical properties of VO$_{2}$ near the
metal-insulator transition.\cite{VO2} In this model, the effective
value of $\varepsilon = \varepsilon _{\mathrm{eff}} (\zeta )$ is
determined by the concentration $f$ ($0 \le f \le 1$) of the metal
phase following the equation,
\begin{equation}
f \cdot \frac{\varepsilon _m - \varepsilon _{\mathrm{eff}}
}{\varepsilon _m + (d - 1)\varepsilon _{\mathrm{eff}} } + (1 -
f)\cdot\frac{\varepsilon _i - \varepsilon _{\mathrm{eff}}
}{\varepsilon _i + (d - 1)\varepsilon _{\mathrm{eff}} } = 0\;,
\end{equation}
where $ \varepsilon _m $ and $\varepsilon _i $ are the
frequency-dependent permittivities for the metallic and insulating
phases, respectively. Also, $d= 2$  and $d = 3$ for the thin film
(thickness smaller then the grain size) and bulk sample,
respectively. In the intermediate region, the effective
permittivity $\varepsilon _{\mathrm{eff}} (i\zeta )$ as a function
of the phase concentration $f$ can be written  as follows:
\begin{eqnarray}
 2\frac{\varepsilon _{\mathrm{eff}} (i\zeta )}{\varepsilon _\infty } = (2f -
1)\frac{\omega_p^2 }{\zeta (\zeta + \gamma )} \\
\nonumber
 + \sqrt {4 + \frac{4\omega_p^2 }{\zeta (\zeta + \gamma )} + \left[
{\frac{(2f - 1)\omega_p^2 }{\zeta (\zeta + \gamma )}}
\right]^2}\,\, {\rm for}\,\, d=2\,,
\end{eqnarray}
and
\begin{eqnarray}
 4\frac{\varepsilon _{\mathrm{eff}} (i\zeta )}{\varepsilon _\infty } = 1 + (3f -
1)\frac{\omega_p^2 }{\zeta (\zeta + \gamma )} + \\
\nonumber
 + \sqrt {9 + \frac{6(1 + f)\omega_p^2 }{\zeta (\zeta + \gamma )} + \left[
{\frac{(3f - 1)\omega_p^2 }{\zeta (\zeta + \gamma )}}
\right]^2}\,\, {\rm for}\,\, d=3\,.
\end{eqnarray}

These equations predict an infinite value of $\varepsilon
_{\mathrm{eff}} (i\zeta )$ when $\zeta \to 0$ (that corresponds to
a metallic conductivity) when $f_c \le f \le 1$ only, where $f_c =
1/d$ is a percolation threshold, see Fig.~\ref{fig3ab}. Otherwise,
a dielectric behavior is present, with a finite value of
$\varepsilon _{\mathrm{eff}} (i\zeta )$ when $\zeta \to 0,$
$$
\varepsilon _{\mathrm{eff}} (\zeta = 0) = \frac{\varepsilon
_\infty}{1-fd} > \varepsilon _\infty , \quad {\rm for} \quad
f<\frac{1}{d}\,,
$$
as shown in Fig.~\ref{fig3ab}. In the metallic region (above the
percolation threshold, for $f
> f_c )$, the behavior of $\varepsilon
_{\mathrm{eff}} (i\zeta )$ at small $\zeta $ is determined by the
effective plasma frequency $\omega_{p, \, \mathrm{eff}} $,
$$\varepsilon _{\mathrm{eff}} \to \varepsilon _\infty \frac{\omega
_{p, \, \mathrm{eff}}^2 }{\zeta (\zeta + \gamma ) } \quad
\mathrm{when} \ \zeta \to 0.$$ The value of $\omega_{p, \,
\mathrm{eff}}^2 $ increases linearly with $f$ from zero, at $f =
f_c $, until $\omega_p^2$, at $f=1$. Thus, a square root behavior
of the effective plasma frequency $\omega_{p, \, \mathrm{eff}} $
over $(f - f_c)$ is present in the metallic region, see inset in
Fig.~\ref{fig3ab}.

It is useful to note here that a linear temperature dependence of
$\omega_p^2$ was observed~\cite{PhaseSep} in
La$_{0.7}$Sr$_{0.3}$MnO$_3$ for $T < T_c $. Thus, we can describe
the Casimir force in the intermediate region as a series of
curves, with their shape only depending on $l / \lambda_p$, as
shown in Fig.~\ref{fig4}.

\section{Predictions for specific materials}
Using the results obtained in the previous sections, here we
estimate numbers for different materials showing the
metal-insulator transition. To study the Casimir force in the
vicinity of the metal-insulator transition, we choose two typical
compounds: vanadium dioxide VO$_{2}$ and the manganites exhibiting
colossal magnetoresistance. For these two materials, the general
Drude behavior of permittivity, with typical values of $\lambda_p$
of the order of 1 $\mu $m and with relatively large values of
$\varepsilon $, $\varepsilon _\infty \sim $ 5--10, is observed in
the  infrared region of interest.

\subsection{Vanadium dioxide VO$_{2}$}

Vanadium dioxide, VO$_{2}$, shows a jump in the static
conductivity (a metal-insulator transition) a little bit above
room temperature, at $T = T_c \approx 68 \ ^\circ$C. The pure
metallic phase of VO$_{2}$ is realized at $T > 88 \ ^\circ $C, and
pure insulator phase~\cite{MIT} (more exactly, semiconducting
phase with a gap of the order of 1~eV) at $T < 60 \ ^ \circ $C.
For vanadium dioxide, the phase separated state has been
observed~\cite{VO2} within a finite temperature range, between 60
$^\circ$C and 88~$^\circ $C, by measuring the optical properties
of VO$_2$. Recently, such state was directly
observed\cite{DirCoexist} via scanning tunneling spectroscopy. For
all temperatures where the metallic conductivity is present, the
generalized Drude behavior is observed up to the infrared
frequency, with a relatively large value of $\varepsilon _\infty
\cong 9$ and $\lambda_p = c / \omega_p \cong 1\ \mu$m. The phonon
contribution to the value of $\varepsilon $, typical for the
infrared region, is screened by free carriers, and the value of
$\varepsilon _\infty \cong 9$ is kept until the high-frequency
region, with $\lambda > \lambda _0 \sim 0.1 \ \mu $m, where the
value of $(\varepsilon - 1)$ vanishes. The value of the
dissipation rate $\gamma $ for this compound is large enough,
$\gamma / \omega_p \sim 0.3$ for VO$_{2}$, and the data for large
$\alpha $ should be considered.

\begin{figure}
 \subfigure[]{\includegraphics[width=\figwidth,height=8cm]{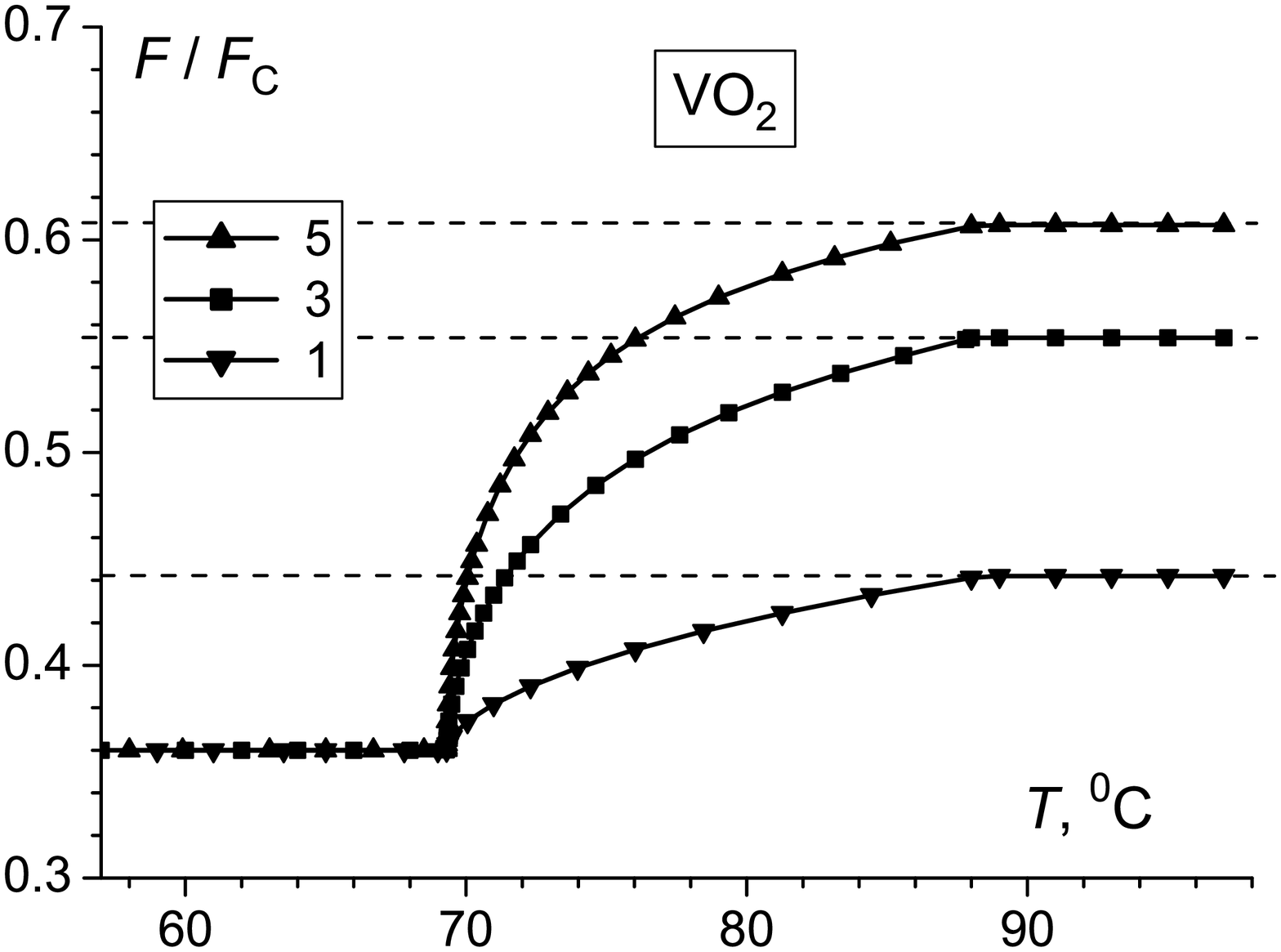}
 \label{zz}}
 \subfigure[]{\includegraphics[width=\figwidth,height=8cm]{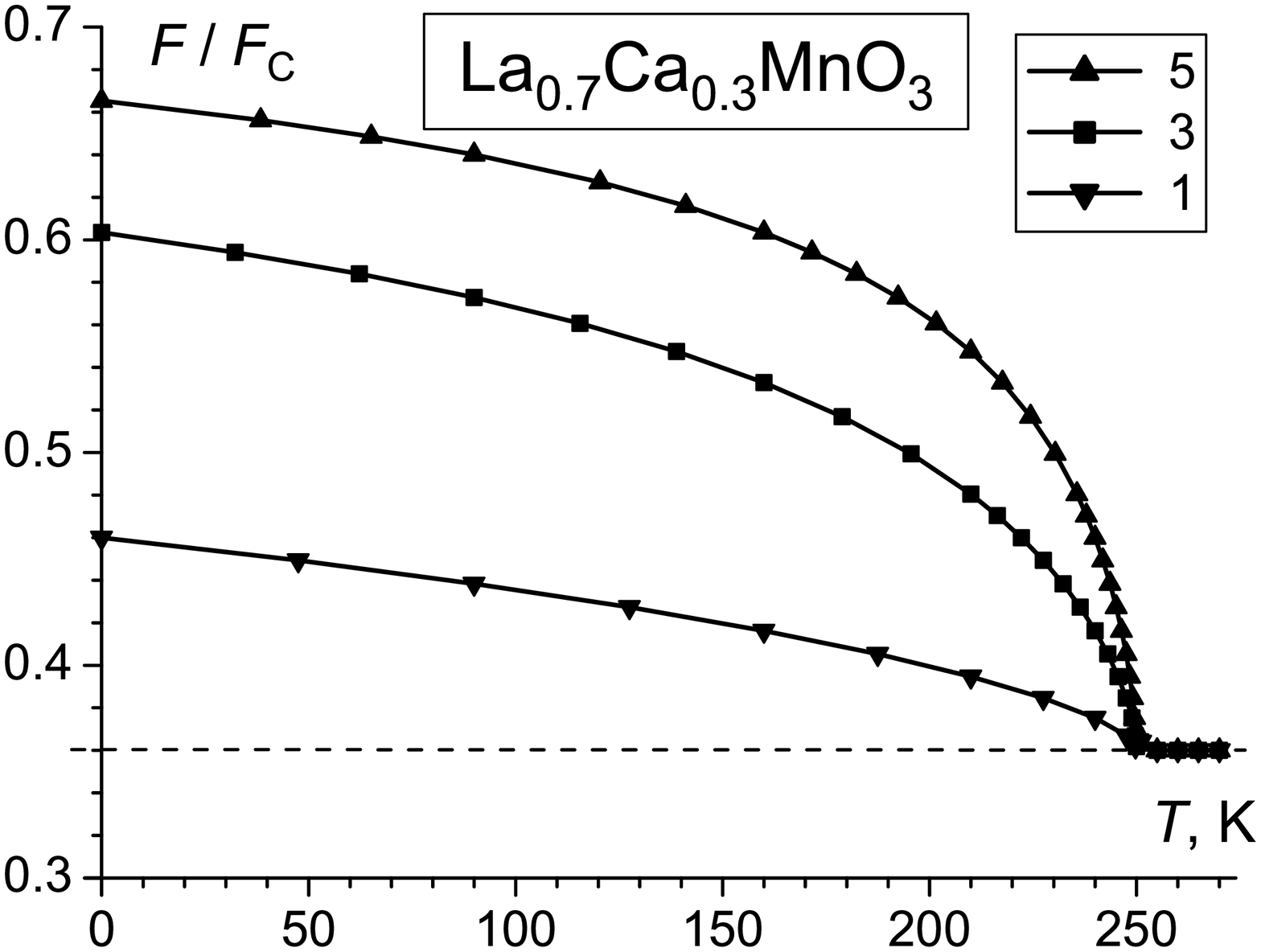}
 \label{yy}}
\caption{\label{Real} Predicted temperature dependence of the
Casimir force for the interaction between  a poor metal and an
ideal metal, calculated for VO$_{2}$ (a) and for
La$_{0.7}$Ca$_{0.3}$MnO$_{3}$ (b). The force is normalized by its
value $F_\mathrm{C}$ for an ideal metal, for different values of
$l / \lambda_p $=1, 3, 5, where $\lambda _p$ is determined by the
plasma frequency in the pure metallic phase. The horizontal dashed
lines indicate the limit values of the force for the pure
insulating phase and (for VO$_{2}$ only) its metallic phase. The
corresponding values of $F_{\mathrm C} $ are described by
Eq.~(\ref{eq1}); for $\lambda_p \simeq 1.2\ \mu$m these are
$F_{\mathrm C} = 0.6 \cdot 10^{- 3}$, $0.74 \cdot 10^{-5}$, and
$0.965 \cdot 10^{-6}$ Dyn/cm$^{2}$ for $l / \lambda_p $=1, 3, 5,
respectively. } \label{fig5}
\end{figure}

For VO$_2$, the Casimir force increases when increasing the
temperature through the transition region, from 60~$^\circ$C until
88 $^\circ$C, see Fig.~\ref{Real}(a). The value of $\alpha $ is
quite high, and the calculated change of the Casimir force is
essentially smaller than for the naive estimate as the difference
between the values $F_{\mathrm{C}}$ for an ideal metal and
$F_{\mathrm{L}}$ for a dielectric, see Eqs.~\eqref{eq1} and
\eqref{eq2}. The relative change of the Casimir force is maximal
for large enough distances, e.g., $l \simeq 4\lambda_p
\simeq$~$\mu$m.

\subsection{Manganites.}
Manganites (with antiferromagnetic insulators LaMnO$_{3}$ or
NdMnO$_{3}$ as parent compounds, after substitution of La by
divalent ion) show a metal-insulator transition at the dopant
concentration $x\sim 0.3$, with a ferromagnetic metallic phase in
the low temperature range.~\cite{CMR} These systems are very
popular now in the context of colossal magnetoresistance, based on
the possibility of the metal-insulator transition induced by an
external magnetic field, that is caused by the ferromagnetic
ordering of the metallic phase. On the other hand, the standard
temperature-induced metal-insulator transition is possible for
such materials as well. For example, the typical compound
La$_{0.7}$Ca$_{0.3}$MnO$_{3}$ demonstrates a metal-insulator
transition at $T = T_c $ = 250 K. The phase separation state is
present for all temperatures below the transition point, and the
typical linear dependence of $\omega_{p, \, \mathrm{eff}}^2 $ has
been observed~\cite{EffMedia} in this region. Note that this
metal-insulator transition is accompanied by ferromagnetic
ordering. In principle, it could produce an extra-force of
magnetic origin near the transition (antiferromagnetic ordering,
present for some metal-insulator transition, does not produce any
source of long-ranged interactions). However, for large enough
plane-parallel samples, the magnetic flux lines are closed inside
the magnetic sample, and should not produce any serious parasitic
effects. For these compounds, $\omega_p$ is small and the
corresponding $\lambda_p \sim
 1 \ \mu$m. The main specific feature important for us here is
the low value of the dissipation rate: typical values of $\gamma /
\omega_p$ are $\sim $ 0.02--0.05, and the low-$\gamma$ behavior of
the curves shown in Figs.~\ref{fig2},~3 are adequate.

For La$_{0.7}$Ca$_{0.3}$MnO$_{3}$,  the metallic phase corresponds
to the low-temperature range, and  the value of the force
increases when decreasing the temperature, which leads to an
opposite temperature behavior of the Casimir force, compared to
VO$_2$. The value of $\alpha $ for this compound is relatively
low, and the temperature dependence of the Casimir force is
sharper than for the previous example. One more specific feature
is the presence of phase separation in the whole region of the
metallic phase existence. Thus, one can expect an essential
dependence of the Casimir force for all temperatures below the
transition temperature, see Fig.~\ref{Real}(b).

\section{Conclusions}
The Casimir force depends on the materials used, and we have
studied some of these material-dependent aspects. The Casimir
force $F_\mathrm{C}$ for a mechanical system containing compounds
with a metal-insulator transition shows an \emph{abrupt}
temperature dependence in the transition region. The relative
change of the force $\Delta F_\mathrm{C} $ when crossing the
transition region can be quite large, of the order of the force
itself for a distance $\sim$ 5--6 microns. The relative change
$\Delta F_\mathrm{C} $ of the Casimir force is larger for large
distances, where the absolute value of the force is small, but
even for a distance $l = \lambda_p = 1.2\ \mu $m it reaches
30{\%}. The dependence of the force on temperature is sharp near
the percolation threshold, where the static metallic conductivity
appears.

When measuring such tiny  forces, the exclusion of any parasitic
effects, like electrostatic forces, is essential. To avoid
electrostatic forces, the usual highly-conducting samples are
short-circuited.~\cite{11} This method might appear to be
ineffective for the metal-insulator transition compounds near the
insulating region. However, such compounds are more semiconducting
than insulating in this region and the conductivity is non-zero at
room temperatures. Thus, we believe that the same technique could
be used. To increase the conductivity in the semiconducting
region, the usual doping by donor or acceptor impurities could be
used. Finally, we note that the metal-insulator transition is
sometimes accompanied by structural phase transitions, which could
lead to some lattice distortions. Thus, care should be taken to
choose materials and operating conditions that avoid these
additional difficulties.

For measurement of the Casimir force for samples made with usual
metals, small separations are preferable. The creation of
experimental set-ups with very small (sub-micrometer) distances
between samples is a serious challenge for experimentalists. As
follows from our analysis, distances $l $ larger than the plasma
wavelength $\lambda_p$ are preferable for the experimental
observation of the effects, we predict around the region of the
metal-insulator transition. For the compounds discussed above,
this means distances of the order of (2--4) $\mu$m. In the planned
experiments~\cite{Exper07} for measuring the Casimir force using
Vanadium oxide samples, the separations  are (0.2--0.4) $\mu$m,
which equals (0.1--0.3) $\lambda_p$. These values are much smaller
than the optimal values noted above. For separations $l$ of the
order of  (0.1--0.3) $\lambda_p$, the Casimir force should follow
low-$l$ asymptotics for any temperature (in both the metallic and
the insulating phases). The temperature dependence of the Casimir
force should be weak and the manifestation of the metal-insulator
transition should be minor for such an experimental set-up.

\begin{acknowledgments}

We gratefully acknowledge partial support from the National Security
Agency (NSA), Laboratory of Physical Sciences (LPS), Army Research
Office (ARO), National Science Foundation (NSF) grant No.
EIA-0130383, JSPS-RFBR 06-02-91200, and Core-to-Core (CTC) program
supported by Japan Society for Promotion of Science (JSPS). S.S.
acknowledges support from the Ministry of Science, Culture and Sport
of Japan via the Grant-in Aid for Young Scientists No 18740224, the
EPSRC via No. EP/D072581/1, EP/F005482/1, and ESF network-programme
``Arrays of Quantum Dots and Josephson Junctions''.

\end{acknowledgments}

\end{document}